\documentclass[twocolumn,amsmath,amssymb,aps,prl]{revtex4}
\usepackage{graphics}
\usepackage{epsfig}

\newcommand{\Pp}{P^{\prime}}
\newcommand{\avgr}{\langle r\rangle}
\newcommand{\avgu}{\langle u\rangle}
\newcommand{\avgz}{\langle z\rangle}

\newcommand{\avgA}{\langle A\rangle}

\newcommand{\medA}{\hat{A}}
\newcommand{\medb}{\hat{b}}

\begin{document}

\title{Scaling in Ordered and Critical Random Boolean Networks}

\author{J.~E.~S.~Socolar}
 \altaffiliation[\vspace{-0.3in}Permanent address: ]{Physics Dept., Duke University, Durham, NC, 27514.}
\author{S.~A.~Kauffman}
\affiliation{Bios Group and Santa Fe Institute, Santa Fe, New Mexico, 87501}

\date{\today}

\begin{abstract}
  Random Boolean networks, originally invented as models of genetic
  regulatory networks, are simple models for a broad class of complex
  systems that show rich dynamical structures.  From a biological
  perspective, the most interesting networks lie at or near a critical
  point in parameter space that divides ``ordered'' from ``chaotic''
  attractor dynamics.  In the ``ordered'' regime, we show rigorously
  that the average number of relevant nodes (the ones that determine
  the attractor dynamics) remains constant with increasing system size
  $N$.  For critical networks, our analysis and numerical results show
  that the number of relevant nodes scales like $N^{1/3}$. Numerical
  experiments also show that the median number of attractors in
  critical networks grows faster than linearly with $N$.  The
  calculations explain why the correct asymptotic scaling is observed
  only for very large $N$.
\end{abstract}
\pacs{87.10.+e,05.45.-a,89.75.-k}

\maketitle

\begin{figure}
  \epsfxsize=0.8\linewidth \epsfbox{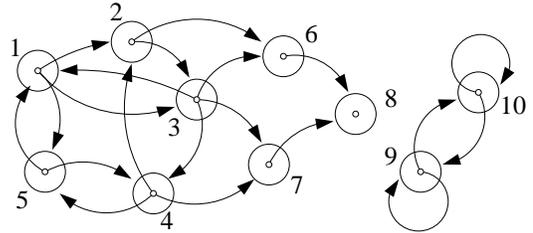} \vspace{-0.1in}
\caption{\label{fig:rbn}
  A $K=2$, $N=10$ network.  Each node has two inputs, but the number
  of outputs (drawn from its center) can vary.  See text for details.}
\end{figure}

A random Boolean network (RBN) is a collection of $N$ binary logic
gates, or nodes, wired together in a random fashion, with each node
implementing a randomly chosen logical function of its inputs.  RBNs
are paradigms for systems in which excitatory and inhibitory
interactions occur among a large set of interacting elements.  One
example of great current interest is the regulatory network that
governs gene expression in a cell.  It has been suggested that the
distinct dynamical attractors of a single RBN be interpreted as
distinct cell types carrying the same genetic information.
\cite{kauffman}
Surprisingly, RBN attractors can exhibit many features of biological
cells, including stability against random external perturbations,
qualitative change in response to special perturbations, and plausible
scaling laws for numbers of attractors and attractor cycle lengths.
\cite{origins} It therefore appears important to understand the
behavior of RBNs as a first step in determining relevant global
properties that might be probed in real gene expression experiments,

Even very simply constructed RBNs with deterministic updating rules
can exhibit a rich set of dynamical behaviors.  We focus here on the
case in which each node has the same number of inputs, $K$.
Fig.~\ref{fig:rbn} shows an example with $K=2$.  Each node $i$
implements a truth function $F_i$ (e.g. AND, XOR, etc.  for K=2) that
is chosen at random from a weighted distribution of all of the
$2^{2^K}$ possible truth functions on $K$ binary inputs.  On each
discrete time step, the outputs are updated synchronously.  Since the
number of states of the system is finite (equal to $2^N$) and the
system is deterministic, for any initial condition the network must
eventually settle into a periodic attractor.  We are interested in the
behavior of large $N$ networks.  How many attractors do they have?
How many nodes typically participate in the attractor dynamics?  

It is well known that tuning the probabilities of different $F$'s can
produce an order-chaos transition. (See \cite{coppersmith} for a
thorough review.)  In the ordered regime, almost all nodes are frozen
and attractor cycles are short.  In the chaotic regime, on the other
hand, the number of fluctuating nodes is a finite fraction of $N$ and
attractor cycles can be quite long.  For large $N$, there is a narrow
critical regime between these phases.  The networks of greatest
interest for biological systems are conjectured to lie near the
critical regime, on the ordered side.  \cite{origins}

In this Letter, we present numerical and analytic results that clarify
the dynamical structure of RBNs in the ordered and critical regimes,
revealing several surprising features: (1) for RBNs in the ordered
regime the average number of relevant nodes (defined below) remains
finite for $N\rightarrow\infty$ and they are organized into trivial
loops; (2) in critical RBNs, the average number of fluctuating nodes
grows like $N^{2/3}$; (3) the system sizes required to observe the
asymptotic regime can be extremely large, especially for $K=2$; and
(4) the median number of attractors in critical RBNs grows faster than
linearly with $N$, at least for $N$ up to 1200.  Both (2) and (4)
contradict previous claims (\cite{parisi} and \cite{bilke},
respectively) which we believe to have been based on studies that did
not consider sufficiently large $N$.  (4) also supersedes an old claim
by one of us that the median number of attractors grows like
$\sqrt{N}$ in critical networks. \cite{kauffman}


The concepts of ``relevant'' nodes \cite{flyvbjerg2} and ``canalizing
inputs'' \cite{kauffman} are essential to our analysis.  In any given
network, there may be nodes whose outputs are frozen at the same value
on every attractor.  Such nodes serve only to fix inputs to other
nodes and are otherwise ``irrelevant''.  There may also be nodes whose
outputs go only to irrelevant nodes.  These are also classified as
irrelevant.  Though they may fluctuate, they act merely as slaves
to the nodes that determine the attractor cycle.

Almost all the irrelevant nodes can be found as follows.  \cite{bilke}
One first identifies ``fixed'' nodes whose outputs are entirely
independent of their inputs.  One then uses an iterative procedure to
identify nodes that must be frozen because their outputs depend only
on inputs from other frozen nodes.  We call all frozen nodes
identified this way ``clamped'', let $s$ denote their number, and
define $u\equiv N-s$.
A similar iterative procedure is then used to remove (or ``prune'')
nodes with no relevant outputs.  For example, in Fig.~\ref{fig:rbn},
even if nodes $1-5$ are all unclamped, nodes $6-8$ can be pruned.  For
the purposes of this paper we designate all nodes that are neither
clamped nor pruned {\em via the described procedure} as ``relevant'',
and let $r$ denote their number.  (Additional nodes may be frozen due
to correlations between two or more unclamped inputs \cite{bilke}, so
$r$ is greater than or equal to the number of truly relevant nodes.)

A canalizing input to a given node is one that can be set to a value
that determines the output, independent of the other inputs.  (For
example, if either input to an OR function is set to 1, the output is
determined.)  Following \cite{flyvbjerg2}, we define a set of
parameters $p_k$ as follows.  For a randomly selected $F$, fix a
randomly selected $K-k$ inputs at arbitrarily chosen values.  $p_k$ is
the probability that those input values are {\em collectively}
canalizing; i.e., that $F$ is independent of the $k$ remaining inputs.
Note that $p_0=1$ (fixing all the inputs certainly determines the
output) and $p_K$ is the probability that a node is fixed (i.e., that
its output is independent of all of its inputs).

The order-chaos transition can be observed by tuning the $p_k$'s.  One
simple way to do this is to assign to each $F$ a probability that
depends only on the number of $1$ in the output column of its truth
table.  For $p\in[0,1]$, we let the probability that a node has truth
function $F$ be $p^{k_1}q^{K-k_1}$, where $q=(1-p)$.  It is
straightforward to check that this parametrization corresponds to
$p_k=p^{2^k}+q^{2^k}$.  With this weighting of the $F$'s, the
transition occurs at $2Kpq=1$, with $2Kpq<1$ corresponding to the
ordered regime \cite{derrida,derridapomeau,derridaweisbuch}.  For our
numerical studies, we set $K=2$ and vary $p$.

\begin{figure}
  \epsfxsize=1.0\linewidth \epsfbox{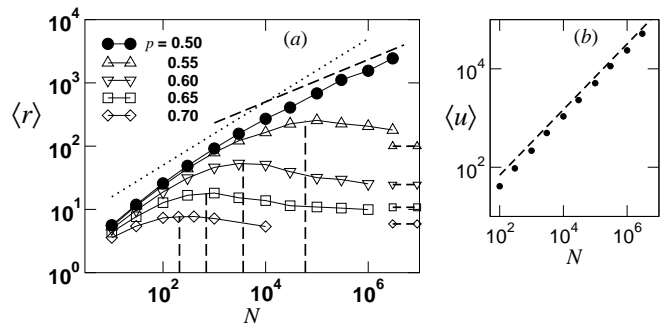} \vspace{-0.2in}
\caption{\label{fig:ruscaling}
  Average numbers of relevant nodes in $K=2$ networks with truth
  functions specified by $p$.  Vertical dashed lines indicate
  theoretical predictions for the crossover from critical to ordered
  behavior at each $p$.  Horizontal segments at the right indicate
  theoretical calculations of the asymptotic values of $\avgr$ for
  each $p$.  Diagonal lines are guides to the eye, showing slopes of
  $1/2$ and $1/3$. (b) Average numbers of unclamped nodes in $K=2$
  critical ($p=1/2$) networks.  The dashed line has slope $2/3$.}
\end{figure}

We have carried out two types of numerical experiment on $K=2$
networks.  First, for 1000 networks at each of five $p$ values, we
determine $u$ and $r$.  Fig.~\ref{fig:ruscaling}(a) shows the measured
$\avgr$ for $N$ up to $3\times 10^6$.  It appears that $\avgr$
approaches a constant at large $N$ for $p$ in the ordered regime.  At
the critical value $p=1/2$, the data are inconclusive, showing
significant curvature on the log-log plot out to the largest $N$ we
have studied.  They are consistent, however, with an asymptotic
scaling law $\avgr\sim N^{1/3}$.  Panel (b) shows the $\avgu$ for
$p=1/2$, indicating a clearer power-law scaling $\avgu\sim N^{2/3}$.

Second, for at least 1000 networks at each $p$, we attempt to measure
the number of attractors, $A$, on the set of relevant nodes.  In some
networks, however, $A$ is prohibitively large, making measurements of
$\avgA$ difficult.  It is much easier to measure the median, $\medA$,
since one need not continue to count attractors in a given network
after the count has exceeded the median.  To count attractors, we
repeatedly choose random initial conditions and identify the attractor
reached.  If 1000 consecutive attempts yield no new attractor, we
record the number of attractors found and move on to another network.
This gives a lower bound on $A$ for each network, and hence a lower
bound on $\medA$.

\begin{figure}[b]
\epsfxsize=0.8\linewidth
\epsfbox{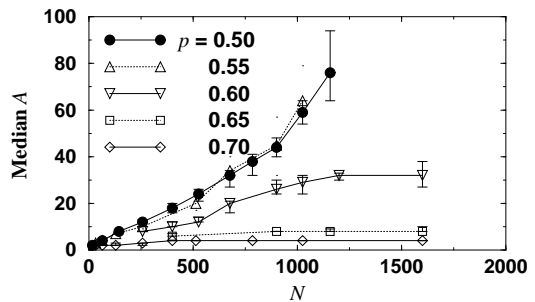}
\vspace{-0.2in}
\caption{\label{fig:attractors} 
  Median numbers of attractors for $K=2$ networks with truth functions
  specified by $p$.  Error bars indicate one standard deviation.  Note
  the upward curvature for the critical value $p=1/2$.}
\end{figure}

Fig.~\ref{fig:attractors} shows the results for $N$ up to 1200.  We
note that in \cite{bilke}, where measurements of $\avgA$ were sought,
it was not possible to consider nets larger than $N=144$.  From
Fig.~\ref{fig:attractors}, however, it is clear that any extrapolation
based on data for $N$ smaller than about 500 at the critical point is
suspect.  We see a faster than linear rise in the median $A$ above
$N\sim 500$, which almost certainly implies a faster than linear rise
in $\avgA$ as well.  Moreover, Fig.~\ref{fig:ruscaling}(a) strongly
suggests that one must study $N>10^6$ to observe the true asymptotic
behavior!

A simple calculation provides a rigorous upper bound on $\avgu$ and
explains why asymptotic scaling sets in only for very large $N$.  A
technical note: in the networks studied numerically, gates were not
permitted to have two inputs from the same node.  The analysis below
ignores this constraint, which yields ${\cal O}(1/N)$ effects.  In all
cases, self inputs are allowed.

Let $P(u)$ be the probability that a randomly selected network has $u$
unclamped nodes.  To compute $P(u)$, we should count the networks with
$u$ unclamped nodes and divide by the total number of networks
$T=[\tau N^K]^N$, where $\tau=2^{2^K}$ is the number of possible truth
functions for each node.  We begin by considering a quantity $\Pp (u)$
that is guaranteed to exceed $P(u)$:
\begin{eqnarray} \label{eq:pp}
\Pp (u) = \frac{1}{T} & C(N,u) &
\left[\sum_{k=0}^{K} 
C(K,k) u^{k}s^{K-k} \tau p_k  \right]^{s} \nonumber \\
\ & \times &  
\left[\sum_{k=0}^{K} 
C(K,k) u^{k}s^{K-k} \tau q_k  \right]^{u},
\end{eqnarray}
where $q_k\equiv 1-p_k$ and $C(m,n)$ is the number of combinations of
$n$ objects drawn from $m$.  The factor $C(N,u)$ is the number of ways
of having $u$ unclamped nodes.  The first sum counts the ways that a
node can be clamped, weighted by the probability of its truth
function: the node in question can have up to $k$ unstable inputs as
long as the other $K-k$ are collectively canalizing, which occurs with
probability $p_k$.  Simlarly, the second sum counts the ways a node
can be unclamped.

$\Pp (u)$ overcounts the probability of having $u$ unclamped nodes
because it ignores the constraint that all clamped nodes must be
traceable through a sequence of inputs back to a fixed node.  That is,
the first sum overcounts the number of ways that $s$ clamped nodes can
be wired, as it includes graphs in which a subset of the $s$ nodes
collectively clamp each other without any connection to a fixed node.
For example, suppose that in Fig.~\ref{fig:rbn} nodes $1$ and $4$ are
fixed, node $9$ implements OR, and the output of $10$ is simply equal
to its input from $9$.  Nodes $2$, $3$, and $5-8$, are clearly clamped
through inputs that can be traced back to $1$ and $4$.  The sum then
counts this network as a possible arrangement of the clamped nodes for
$u=0$ because it is self-consistent to assume that both $9$ and $10$
are clamped.  (Note that after a few time steps 9 and 10 will either
both be stuck on 0 or both on 1.)  However the iterative procedure
defined above would (correctly) not identify $9$ and $10$ as clamped,
so its inclusion in $\Pp(0)$ constitutes overcounting.  Note that the
same network is also (properly) counted in $\Pp(2)$.  Thus $\Pp(u) >
P(u)$ for all $u$, implying that $\sum_u g(u)\Pp(u)\geq \sum_u g(u)P(u)$
for any $g(u)\geq 0$.

If $u$ is small compared to $N$, a useful approximation to
Eq.~(\ref{eq:pp}) can be obtained.  Using $s=N-u$ and using Stirling's
formula to simplify the binomial coefficients, together with the
identity $(1+x/n)^n = \exp[x(1-(x/n)/2+(x/n)^2/3+\dots)]$, we find
\begin{equation} \label{eq:ppapprox}
\Pp(u) \simeq \frac{1}{\sqrt{2\pi u(1-u/N)}}
\exp \left[-\theta_1 u - \theta_2 \frac{u^2}{N} - \theta_3 \frac{u^3}{N^2}\right],
\end{equation}
where 
\begin{eqnarray}
\theta_1 & = & Q - 1 -\ln Q, \\
\theta_2 & = & (1-Q)(1+Q-2K+2Q_2)/2, \\
\theta_3 & = & -\theta_2 - P_3 - Q_3 + (K-Q)P_2 + \frac{1}{2}Q_2^2 \nonumber \\
      \  & \ &     - [K-1+2(K-Q)^3]/6,
\end{eqnarray}
with $Q\equiv Kq_1$, $Q_n\equiv C(K,n) q_n/Q$, and $P_n\equiv C(K,n)
p_n$.  The calculation involves only straightforward algebra and the
assumption that terms of order $u^4/N^3$ can be neglected in the
exponent, which is justified whenever $\theta_1$, $\theta_2$, and
$\theta_3$ are all positive and at least one of them is nonzero.
Under such circumstances, and in the limit of large $N$, $\Pp(u)$ is
exponentially strongly suppressed for $u$'s larger than order
$N^{2/3}$, whereas the neglected terms would only become relevant for
$u$'s of order $N^{3/4}$.

$Q$ is equivalent to the order parameter defined by Flyvbjerg in
\cite{flyvbjerg2}, where it was also argued that $Q=1$ marks the
critical boundary.  Eq.~(\ref{eq:ppapprox}) proves that for all $Q<1$
the $P(u)$ decays exponentially with a decay length independent of
$N$.  It also strongly suggests that this is not the case at $Q=1$,
though the computation only gives an upper bound.  The fact that
$\theta_1$ and $\theta_2$ {\em both} vanish at $Q=1$ is a surprising
result that affects the scaling of $\avgr$ in critical networks, as we
shall see below.

In the ordered regime $Q<1$, we have $\theta_1>0$, so higher order
terms are irrelevant at sufficiently large $N$ and $\Pp(u)$ becomes
independent of $N$.  Thus $\avgu$ for asymptotically large $N$ is
bounded above by a constant.  For the case where $p_k$ is determined
simply by the one parameter $p$, we have $q_1 = 2p(1-p)$ and the
critical $p$ satisfies $2Kp_c(1-p_c)=1$, consistent with previous
studies of the effect of $p$.  For $p$ near $p_c$ we get
\begin{equation}
\theta_1 \simeq 
\left\{\begin{array}{ll} 
  8(p-p_c)^4  & \quad {\rm for\ } K=2 \\ \nonumber
  2K(K-2)(p-p_c)^2 & \quad {\rm for\ } K>2.  
\end{array}
\right.
\end{equation}
For $K=2$, then, $\theta_1$ is quite small even for $p$ relatively far
from $p_c=1/2$.  For example, $p=0.55$ gives $\theta_1 \approx 5\times
10^{-5}$.  The asymptotic behavior of the system should only be
apparent when $u$ is bigger than $1/\theta_1$ and hence when $N$ is
substantially larger than that.  The vertical dashed lines in
Fig.~\ref{fig:ruscaling}(a) are drawn at $N=3/\theta_1$, which is seen to
give an accurate indication of where effects associated with the
ordered regime become important.

Moreover, for very large $N$, we may obtain an accurate approximation
to $P(u)$ simply by normalizing $\Pp(u)$, since the overcounting
factor relating the two approaches some nonzero constant at $u=0$.
This allows us to compute $\avgr$ (not just an upper bound on it) as
follows.  Consider the unclamped nodes and the links between them.
These form a random graph subject only to the constraint that each
node has between 1 and $K$ inputs.  But from the second sum in
Eq.~(\ref{eq:pp}) we know the relative probabilities that an unclamped
node will have $k$ unclamped inputs.  Specializing to
$K=2$ for simplicity, the average number of inputs per node for fixed
$u$ is
\begin{eqnarray}
\avgz & = & \left(Q s u + 2 q_2 u^2\right)/\left(Q su + q_2 u^2\right) \nonumber \\
   \    & \simeq & 1 + \frac{q_2}{Q}\frac{u}{N}+\left(\frac{q_2}{Q}-\frac{q_2^2}{Q^2}\right)\frac{u^2}{N^2}. \label{eq:z} 
\end{eqnarray}
Now in the ordered regime, where the exponential cutoff in $P(u)$ is
independent of $N$, we have $\int\!du P(u)u/N\!\rightarrow\!0$ 
for all $u$ as $N\!\rightarrow\!\infty$ and hence $\avgz\rightarrow 1$.
Thus the network of unclamped nodes at large $N$ is just a randomly
wired $K=1$ network with no fixed nodes.  Exact combinatorics for
$K=1$ random graphs show that the expected number of loops of size $n$
in a network of size $u$ is $L_u (n) = u!/[n (u-n)!\,(u-1)^n].$
\cite{flyvbjerg} Since all nodes not in loops can be pruned, we obtain
\begin{equation}
\avgr = \sum_{u=0}^N\sum_{n=1}^{u}n\,L_u(n)\,P(u) \simeq 0.7/\sqrt{\theta_1}.
\end{equation}
The horizontal dashed lines on the far right in
Fig.~\ref{fig:ruscaling}(a) mark this prediction, which agrees well
with the data.

For the critical case $Q=1$, we have
\begin{equation} \label{eq:ppcritical}
\Pp(u) \simeq [2\pi u(1-u/N)]^{-1/2}
\exp \left[- \theta_3 u^3/N^2\right],
\end{equation}
which implies that $\avgu$ cannot grow faster than $N^{2/3}$.  This
contradicts a previous argument suggesting $\avgu\sim N^{3/4}$
\cite{parisi}.  Note however, that the corrections from terms of order
$u^4/N^3$ cannot be neglected unless $(N^{2/3})^4/N^3 = N^{-1/9} <<
1$, suggesting that the critical scaling emerges only for $N>10^9$, as
confirmed by Fig.~\ref{fig:ruscaling}.

Naive normalization of $\Pp(u)$ to get $P(u)$ yields $\avgu\sim
N^{2/3}$ and Fig.~\ref{fig:ruscaling}(b) indicates that this is
correct.  Given the scaling law and the fact that paths connecting
relevant nodes cannot lead to dead ends, the theory of directed random
graphs \cite{newman} and Eq.~(\ref{eq:z}) can be used to show that
$\avgr\sim N^{1/3}$, consistent with the numerical trend seen in
Fig.~\ref{fig:ruscaling}(a).  Unfortunately, naive normalization does
not yield an accurate prediction for the full form of $P(u)$ because
the $u$ dependence of the factor $f(u)\equiv P(u)/\Pp(u)$ becomes
important.  Determination of the precise form of $f(u)$ is beyond the
scope of this work.

We now address the question of the number of attractors.  For the
case $Q<1$ and very large $N$, the relevant nodes form trivial loops
with only two possible truth functions: $F(\sigma)=\sigma$
(``identity'') and $F(\sigma)=1-\sigma$ (``not'').  For $n$ prime, a
loop of size $n$ has either $(2^n-2)/n$ or $(2^n-2)/2n$ attractors,
depending on whether the number of nots in the loop is even or odd.
For $n$ not prime, the number of attractors is slightly larger.
\cite{flyvbjerg} Now $L_u(n)(2^n-2)/n$ is extremely sharply peaked
very close to $n=u/2$.  $\avgA_u$ is thus dominated by rare networks
that contain a relevant loop of size $u/2$, which gives $\avgA_u\sim
2^{cu/2}$, with $c$ a constant of order unity.  Naive averaging over
$P(u)$ gives a divergent answer; a consistent calculation would
require inclusion of the terms in $P(u)$ that cause rapid decay for
$u$ of order $N$.

The median number of attractors, $\medA$, can be estimated as
$2^{\medb}/\medb$, where $b$ is the size of the biggest loop of
relevant nodes in a given network.  For fixed $u$, the probability of
occurence of $b$ is $P_u(b) = L_u(b)
\prod_{n=b+1}^u\left[1-L_u(n)\right]$.  To find $\medb$, we
numerically solve $ \sum_{b=0}^{\medb} \sum_{u=0}^{\infty} P(u) P_u(b)
= 1/2$.  The results for the networks studied in
Fig.~\ref{fig:attractors} are $\medA = 3$, $5$, and $14$ for $p=0.7$,
$0.65$, and $0.6$, respectively.  This result and its rough agreement
with the data (see Fig.~\ref{fig:attractors}) confirms that $\medA$
approaches a constant at large $N$ for any sub-critical $p$
\cite{lynch} and illustrates the qualitative distinction between
$\avgA$ and $\medA$.


In closing, we note that the classification of nodes as relevant is
robust across a large class of updating rules.  The results on $\avgr$
reported here apply to asynchronous models and to models in which
there is a stochastic time delay in the updating of each node, as long
as the update, whenever it occurs, is always accurate.  Finally, we
note that the numbers of genes in real cells are on the order of
$10^4$, which our results show may be too small to exhibit asymptotic
large $N$ behavior.  Thus, networks with canalization parameters that
nominally place them in the ordered regime will exhibit features of
critical networks, which may be important for biological function.

{\it Acknowledgments:} We thank M. Newman, M. Girvan, S. Coppersmith,
R. Palmer, and H. Flyvbjerg for helpful communications.  J.E.S.S.
thanks Bios Group, the Santa Fe Institute, and the Smith Faculty
Enrichment Fund at Duke University for financial and facilities support.
  

\end{document}